\documentclass[twocolumn, tighten]{aastex61}
 
\usepackage{amsmath,amstext}
\usepackage{hyperref}

\shorttitle{Apocenter Pile-Up}
\shortauthors{Deason et al.}

\begin{document}
\title{Apocenter Pile-Up: Origin of the Stellar Halo Density Break }

\correspondingauthor{Alis J. Deason}
\email{alis.j.deason@durham.ac.uk}

\author[0000-0001-6146-2645]{Alis J. Deason}
\affiliation{Institute for Computational Cosmology, Department of Physics, University of Durham, South Road, Durham DH1 3LE, UK}

\author{Vasily Belokurov}
\affiliation{Institute of Astronomy, University of Cambridge, Madingley Road, Cambridge CB3 0HA, UK}
\affiliation{Center for Computational Astrophysics, Flatiron Institute, 162 5th Avenue, 10010, New York, NY, USA}
\author[0000-0003-2644-135X]{Sergey E. Koposov}
\affiliation{McWilliams Center for Cosmology, Department of Physics, Carnegie Mellon University, 5000 Forbes Avenue, Pittsburgh, PA 15213, USA}
\affiliation{Institute of Astronomy, University of Cambridge, Madingley Road, Cambridge CB3 0HA, UK}

\author{Lachlan Lancaster}
\affiliation{Department of Astrophysical Sciences, Princeton University, 4 Ivy Lane, 08544, Princeton, NJ, USA}

\begin{abstract}
We measure the orbital properties of halo stars using 7-dimensional information provided by \textit{Gaia} and the Sloan Digital Sky Survey. A metal-rich population of stars, present in both local main sequence stars and more distant blue horizontal branch stars, have very radial orbits (eccentricity $\sim 0.9$) and apocenters that coincide with the stellar halo ``break radius" at galactocentric distance $r \sim 20$ kpc. Previous work has shown that the stellar halo density falls off much more rapidly beyond this break radius. We argue that the correspondence between the apocenters of high metallicity, high eccentricity stars and the broken density profile is caused by the build-up of stars at the apocenter of a common dwarf progenitor. Although the radially biased stars are likely present down to  metallicities of [Fe/H] $\sim -2$ the increasing dominance at higher metallicities suggests a massive dwarf progenitor, which is at least as massive as the Fornax and Sagittarius dwarf galaxies, and is likely the dominant progenitor of the inner stellar halo.
\end{abstract}


\section{Introduction}
Stars on elliptical orbits move rapidly through their point of closest approach and slow down at their furthest extent. This continual speed-up/slow-down cycle is akin to cars on a highway speeding through the open road, and turning on the brakes at the onset of traffic. Naturally, as cars behave in heavy traffic, the inevitable slow-down leads to a ``pile-up" of stars at apocenter. This phenomenon can lead to striking features in galaxy stellar halos, which are formed from the continual digestion of smaller mass dwarf galaxies. Stars stripped from dwarfs on very radial orbits can appear as shell-type features, which are a build-up of stars at apocenter \citep[e.g.,][]{johnston08, cooper11}. Thanks to the very long dynamical times in the halo, these features can persist over several Gyr and, like their stellar stream counterparts, display a visible memory of the galaxy's accretion history. 

\begin{figure*}
    \centering
    \includegraphics[width=16cm, height=5.33cm]{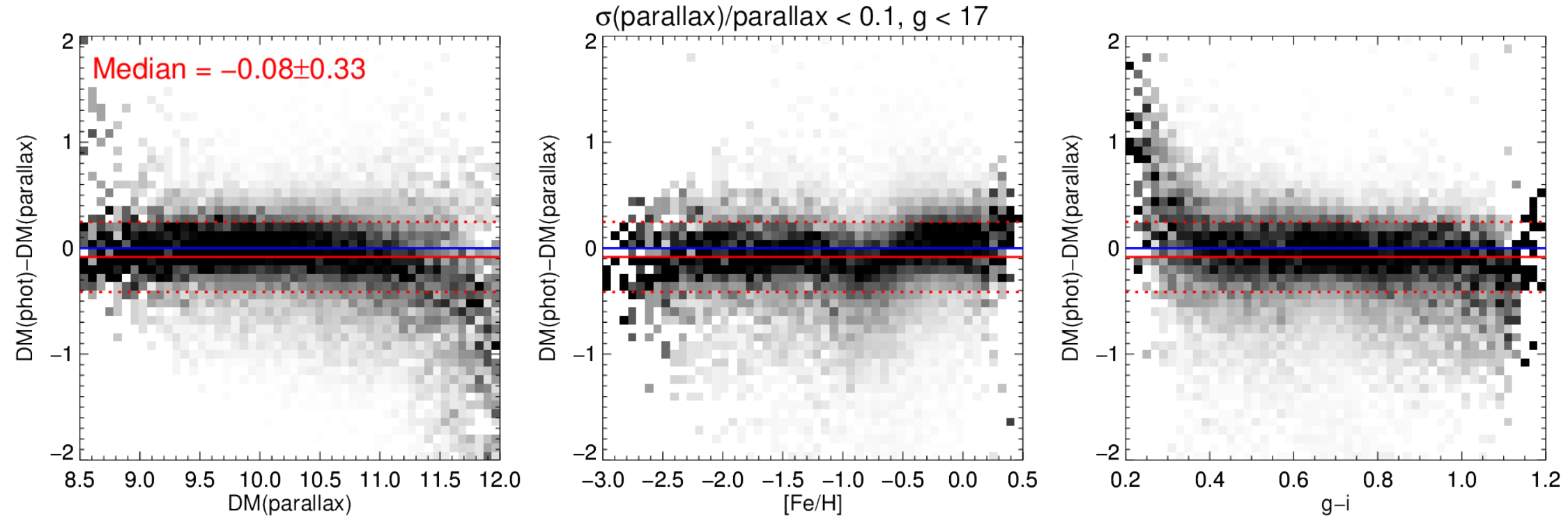}
    \caption{Distance modulus of main sequence stars estimated using photometric parallax \citep{ivezic08} and \textit{Gaia} astrometric parallax. Here we use the distances provided by \cite{cbj18}, and only consider bright stars with accurate parallax measurements ($\leq 10 \%$). We compare the derived distance measurements as a function of distance modulus (left), metallicity (middle) and $g-i$ color (right). Note that the density plots are column normalized. The photometric estimates agree well with the astrometric parallax estimates. There is a small offset ($-0.08$ dex) and 0.33 dex scatter, with little variation with distance, metallicity or color. This comparison shows that the photometric parallax can be used to measure main sequence star distances to $\sim 15\%$ accuracy.}
        \label{fig:abm}
\end{figure*}

In recent years, it has been recognized that the halo star counts in
the Milky Way display a peculiar feature. Namely, instead of following
a simple power-law density distribution, the halo density profile
exhibits a ``break" at galactocentric distances of $r \sim 20$ kpc,
whereby the star counts fall-off much more rapidly beyond the break
radius \citep[e.g.,][]{watkins09, deason11, sesar11, pila15,
  xue15}. While the details of the density profiles vary, these works
find shallower power-law slopes ($\alpha \sim 2.5$) inside the break
radius, and steeper power-laws ($\alpha \sim 3.7-5$) beyond it. In
\cite{deason13} we used a suite of simulated stellar halos
\citep{bullock05} to argue that these broken halo profiles are due to
a build-up of stars at apocenters --- either due to the accretion of a
small group of dwarfs at similar times, or the accretion of one
massive dwarf. Evidence for the latter scenario has rapidly been
growing \citep[e.g.,][]{deason15, fiorentino15, belokurov18a,
  lancaster18}. \cite{belokurov18b} recently showed that $\sim 2/3$ of
the material in the inner stellar halo exhibits extreme radial
anisotropy making it appear ``sausage-like'' in velocity space - this,
they argue, is a consequence of the accretion of a massive dwarf galaxy
on a highly eccentric orbit. Fast on the heels of the second
\textit{Gaia} data release (DR2) \cite{myeong18} found that $N =8$ of
the Milky Way globular clusters are likely related to this markedly
radial accretion event. Indeed, the association of a large number of
globular clusters provides further evidence that the ``sausage'' is
related to a massive halo progenitor. These findings are in good 
agreement with \cite{kruijssen18}, who use the age-metallicity distribution 
of Galactic globular clusters to infer the halo's assembly history.

In this Letter, we use \textit{Gaia} DR2 proper motions to derive the
apocenter and pericenter distributions of halo stars in the Milky
Way. Now with \textit{Gaia} we can, for the first time, relate the
orbital properties of halo stars to the broken density profile feature
that was discovered almost 10 years ago.

\section{Halo Stars in 7D}
We construct samples of halo stars with 6D phase-space and metallicity (``7th dimension") measurements. These comprise a local  ($D \lesssim 5$ kpc) sample of main sequence stars, and a more distant sample of Blue Horizontal Branch (BHB) stars. In both cases, spectroscopic measurements are derived from the Sloan Digital Sky Survey (SDSS), and astrometry is taken from the newly released \textit{Gaia} DR2 catalog \citep{gaia_mission, gaia_dr2}.

\subsection{Local Main Sequence Stars}
\begin{figure}
    \centering
    \includegraphics[width=8.5cm, height=6.4cm]{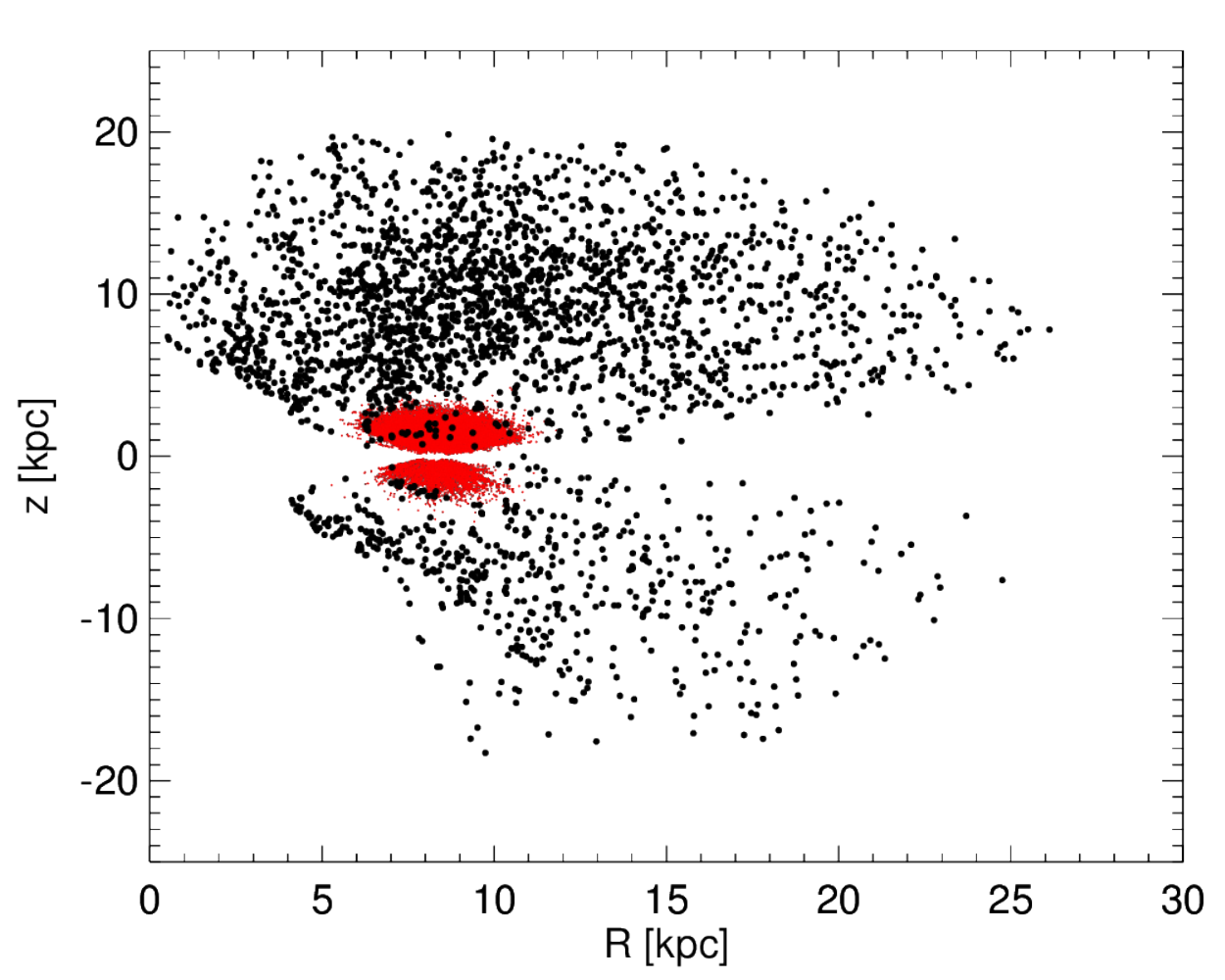}
    \caption{The spatial distribution of the main sequence (red) and blue horizontal branch (black) stars in cylindrical ($R, z$) coordinates. Here, the Sun is located at $(R, z)= (8.3, 0)$ kpc. }
    \label{fig:spatial}
\end{figure}

\begin{figure*}
  \begin{minipage}{\linewidth}
    \centering
    \includegraphics[width=16cm, height=5.33cm]{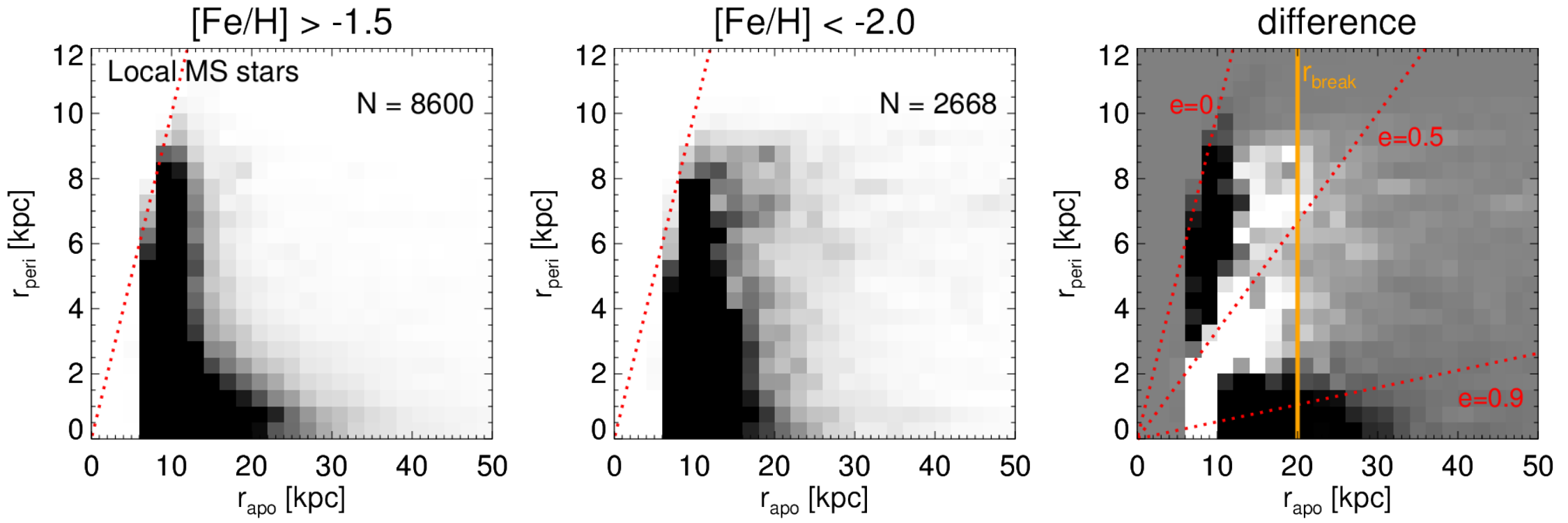}
  \end{minipage}
  \begin{minipage}{\linewidth}
    \centering
    \includegraphics[width=16cm, height=5.33cm]{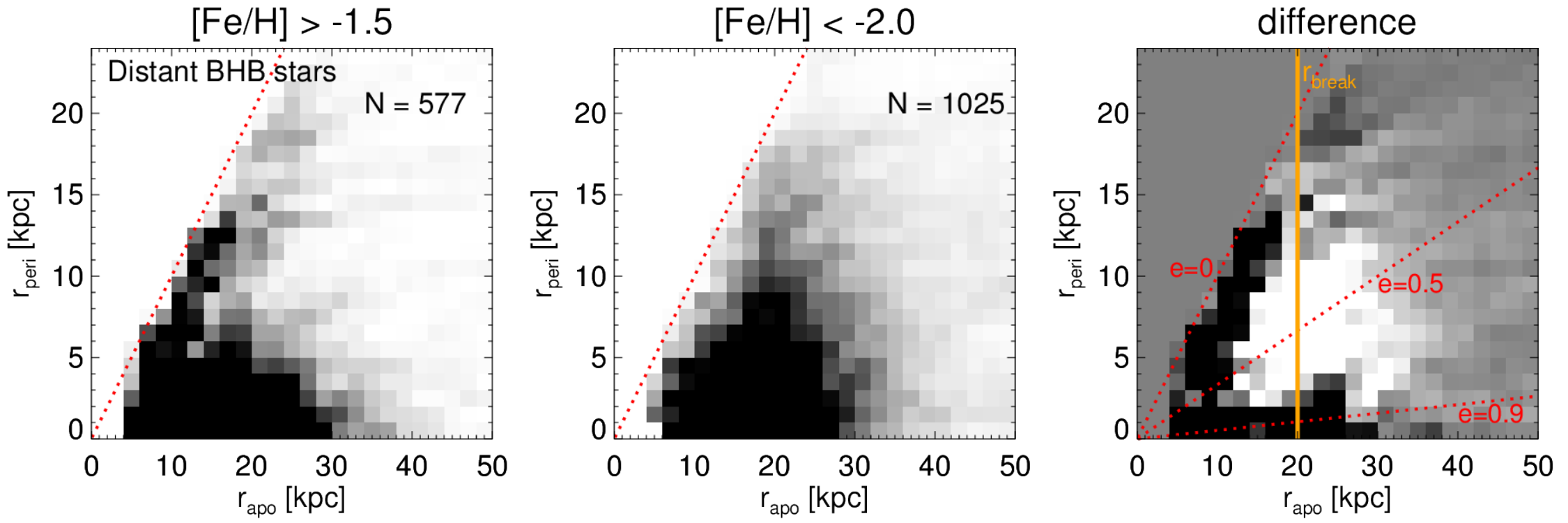}
  \end{minipage}
    \caption{The apocenter and pericenter distributions of the local
      main sequence stars (top panels) and distant BHB stars (bottom
      panels).  We show the 2D distributions for 
      metal-rich ([Fe/H] $>-1.5$) and metal-poor ([Fe/H] $< -2 $) stars in the left and middle
      panels. The metal-rich minus metal-poor difference is shown in
      the right panels. Here, gray indicates no difference, black is
      an excess of metal-rich stars and white is an excess of
      metal-poor stars. Tracks of constant eccentricity ($e=0, 0.5,
      0.9$) are shown with the dotted red lines, and the vertical
      orange lines indicated the approximate break radius of the
      stellar halo \citep{deason11}. In both the local and distant
      samples two clear residuals stand out in the metal-rich stars:
      (1) the disk population with $e \sim 0$ (for the BHB sample, the
      ``disk'' population is likely supplied by a small number of
      contaminating Blue Stragglers), and (2) a population with very
      high eccentricity ($e \sim 0.9$) and apocenters coincident with
      the break radius --- the ``sausage" stars.}
    \label{fig:rperi_apo}
\end{figure*}

We select main sequence stars from the SDSS DR9 spectroscopic catalog \citep{ahn12} by applying the following cuts on color, surface gravity and effective temperature: $0.2 < g -i < 2$, $ 0.2 < g-r < 0.8$, $3.5 < \mathrm{log}(g) < 5$, $4500 < T_{\rm eff}/K < 8000$. We exclude from our sample stars with low signal-to-noise spectra (S/N $<10$), large line-of-sight velocity errors ($\sigma_{RV} > 50$ km s$^{-1}$), and high extinction ($A_g > 0.5$). We also restrict to Galactic latitudes $|b| > 10^\circ$ and relatively low metallicities [Fe/H] $< -1$ to minimize the presence of disk stars in the sample. Finally, we limit our sample to magnitudes $g < 17$ to ensure that we have accurate spectroscopic and astrometric measurements. The sample is cross-matched with the \textit{Gaia} DR2 source catalog, resulting in $N=18,185$ main sequence stars with proper motion measurements.

We estimate the stars' distances using the relations given in
\cite{ivezic08} (equations A2, A3 and A7). In Figure \ref{fig:abm} we
compare these photometric parallaxes to distance estimates based on
astrometric parallaxes from \textit{Gaia}. Instead of simply inverting
the parallax, we use the probabilistically inferred astrometric
distances derived by \cite{cbj18}. Please note that a more
self-consistent, but not immediately available, approach would involve
using the method described in \cite{cbj18} to estimate distance moduli
from Gaia's parallaxes. We find a small offset (-0.08 dex) in distance
modulus between the astrometric and photometric distance estimates,
and a scatter of 0.33 dex. The offset could be due to small biases in
the astrometric parallaxes themselves, so we do not attempt to correct
for this bias. However, it is reassuring that the photometric
parallaxes can be used to measure distances to 15\% with little
dependence on metallicity (middle panel) or color (right panel).

\subsection{Distant Blue Horizontal Branch Stars}
To probe to further distances in the halo, we utilize the SDSS/SEGUE Blue Horizontal Branch (BHB) sample compiled by \cite{xue11}. Here, we consider relatively bright ($g <17$) stars in the sample, which go out to $\sim 20$ kpc. We apply the color and metallicity dependent absolute magnitude relation derived by \cite{fermani13} to estimate distances to the stars. We assume the distance calibration for BHBs is accurate to 5\% \citep[cf.][]{deason11, fermani13}. As this calibration was only applied to stars redder than $g-r > -0.4$, we exclude the (small number of) very blue stars with $g-r < -0.4$. After cross-matching with the \textit{Gaia} source catalog we obtain $N =2,700$ BHB stars with 7D measurements.
\\
\\
\noindent
In Fig. \ref{fig:spatial} we show the spatial distribution of our halo star samples. In what follows we assume a circular velocity of $V_c=235$ km s$^{-1}$ at the position of the Sun ($R_\odot =8.3$ kpc), and solar peculiar motion $(U_\odot, V_\odot, W_\odot) = (11.1, 12.24, 7.25)$ \citep{schonrich10, reid14}.

\section{Orbital Properties}
\begin{figure}
  \begin{minipage}{\linewidth}
    \centering
    \includegraphics[width=8.5cm, height=7.2cm]{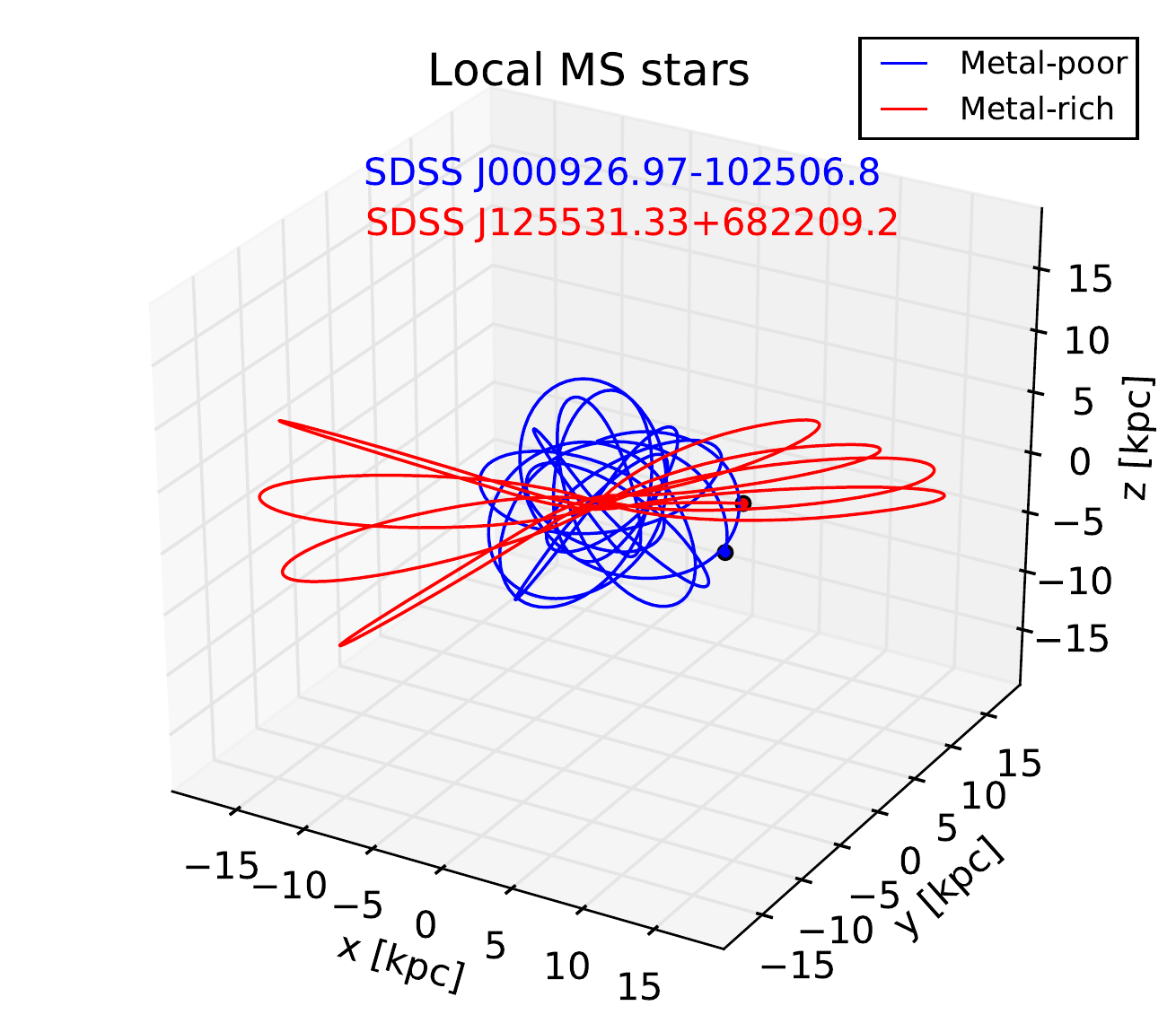}
  \end{minipage}
  \begin{minipage}{\linewidth}
    \centering
    \includegraphics[width=8.5cm, height=7.2cm]{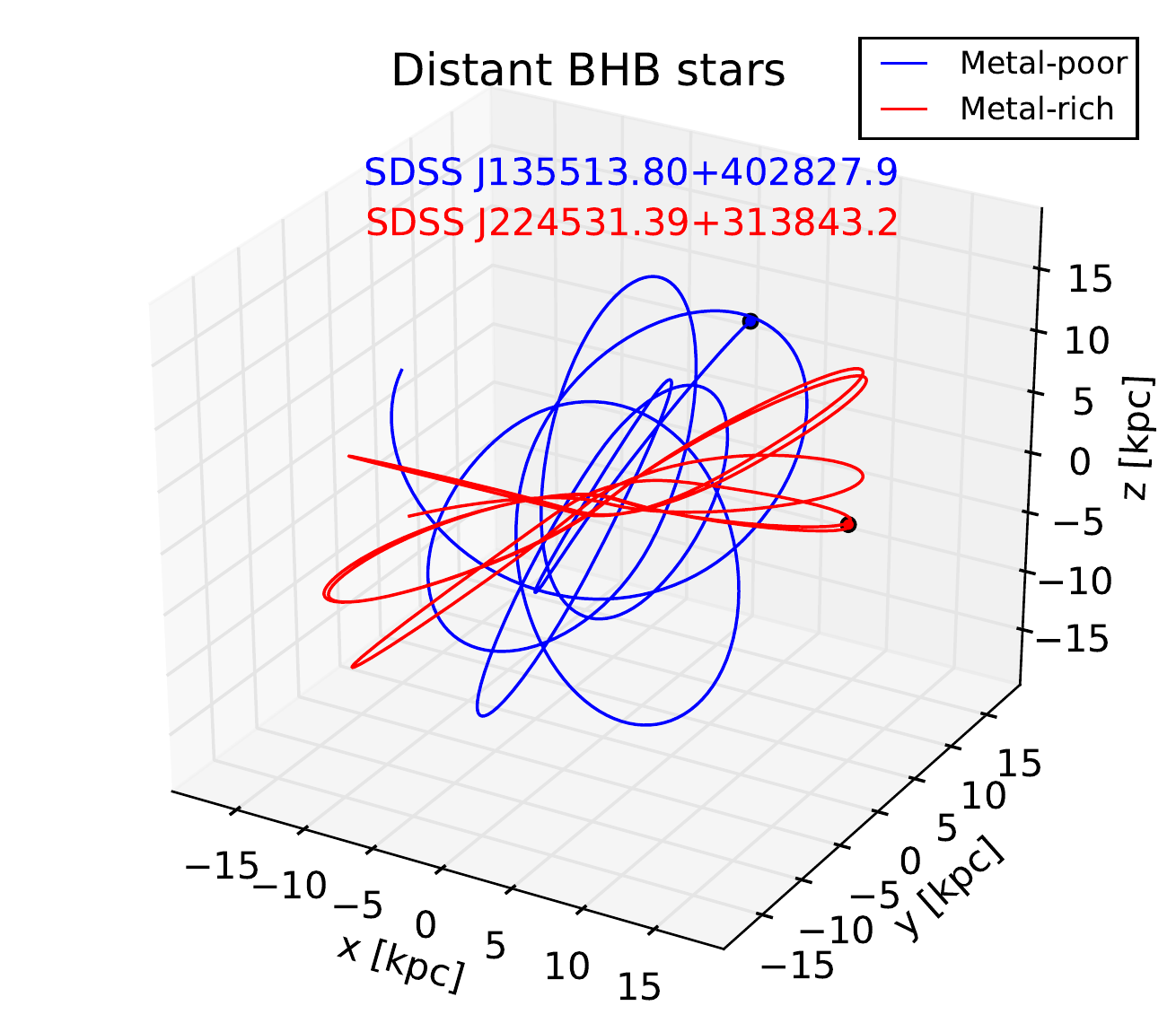}
  \end{minipage}
    \caption{Example orbits in 3D $(x, y, z)$, where the Galactic center is at $(x, y, z) = (0, 0, 0)$. The red lines show example orbits of metal-rich stars on radial orbits, and the blue lines show metal-poor stars on more isotropic orbits.}
    \label{fig:orbits}
\end{figure}

In this Section, we derive the apocenter and pericenter distributions of our 7D halo samples. The orbits are calculated using the \texttt{galpy}\footnote{\url{http://github.com/jobovy/galpy}} software package developed by \cite{bovy15}. We adopt the \texttt{MWPotential2014} gravitational potential, which is described in Table 1 in \cite{bovy15}. However, we find that our results are not significantly changed if we adopt other potentials commonly used in the literature \citep[e.g.][]{mcmillan17}. To propagate errors in proper motion (including covariances), line-of-sight velocity and distance in to orbital parameters we use Monte Carlo sampling. We typically find that the apocenters and pericenters of our halo stars have uncertainties of $\sim0.5$ kpc and $\sim1$ kpc for the local and distant samples, respectively.

In Fig. \ref{fig:rperi_apo} we show the apocenter and pericenter
distributions for the main sequence stars (top panels) and BHB stars
(bottom panels). In the left panels we show metal-rich
stars ([Fe/H] $> -1.5$), the middle panels show metal-poor stars ([Fe/H] $ < -2$), and the right panels
show the difference between the metal-rich and metal-poor
distributions (metal-rich minus metal-poor). Here, black indicates an
excess of metal-rich stars, and white indicates an excess of
metal-poor stars. The comparison between metal-rich and metal-poor
stars clearly shows two residuals in the metal-rich stars: (1) a disk
population with eccentricity $e\sim 0$ and (2) a component with high
eccentricity ($e \sim 0.9$). Remarkably, the latter ``sausage-like"
stars are seen in both the local sample of main sequence stars
\citep[cf.][]{belokurov18b} and in the more distant BHB
sample. Moreover, we find that the apocenters of this population are
coincident with the break radius of the stellar halo. 

The range of apocenters ($10 \lesssim r_{\rm apo}/\mathrm{kpc} \lesssim 30$) of the high eccentricity (``sausage'') stars seen in Fig. \ref{fig:rperi_apo} is related to the spread in energy of
their dwarf progenitor, and the number of orbits since infall. We note that a narrow range of apocenters would 
suggest a very recent and/or relatively low-mass accretion event, which does not appear to be the case here (see below). In \cite{deason13} we show that the
break radii in the Bullock \& Johnston stellar halo simulations
approximately correspond to the \textit{average} apocenter of
the stars stripped from the same progenitor (see Figure 2 of the paper). The 
truncation at $r \sim 25-30$ kpc signifies the outermost apocenter of
the debris, which have the highest energy orbits. 

The average apocenter for the high eccentricity ($ e > 0.9$), metal-rich ([Fe/H] $> -1.5$) stars is 
$16 \pm 6$ kpc and $20 \pm 7$ kpc for the main sequence and BHB stars, respectively. Note, here the 
error bars give the standard deviation about the average. These average apocenters are in 
excellent agreement with measurements of the break radius of the Milky Way stellar halo, and thus
confirm the predictions made by \cite{deason13}. 

We show examples of the orbits of the main sequence and BHB stars in
Fig. \ref{fig:orbits}. Here, we give cases of metal-poor (blue lines)
and metal-rich (red lines) stars. The local high [Fe/H] main sequence
stars on highly radial orbits are very similar to the distant BHB
stars --- they are just at different points in the orbit. Indeed, the
common apocenters shared between the more distant sample and the local
sample suggest that they originate from the same progenitor.

\begin{figure*}
  \begin{minipage}{0.54\linewidth}
    \centering
   \includegraphics[width=9cm, height=7.2cm]{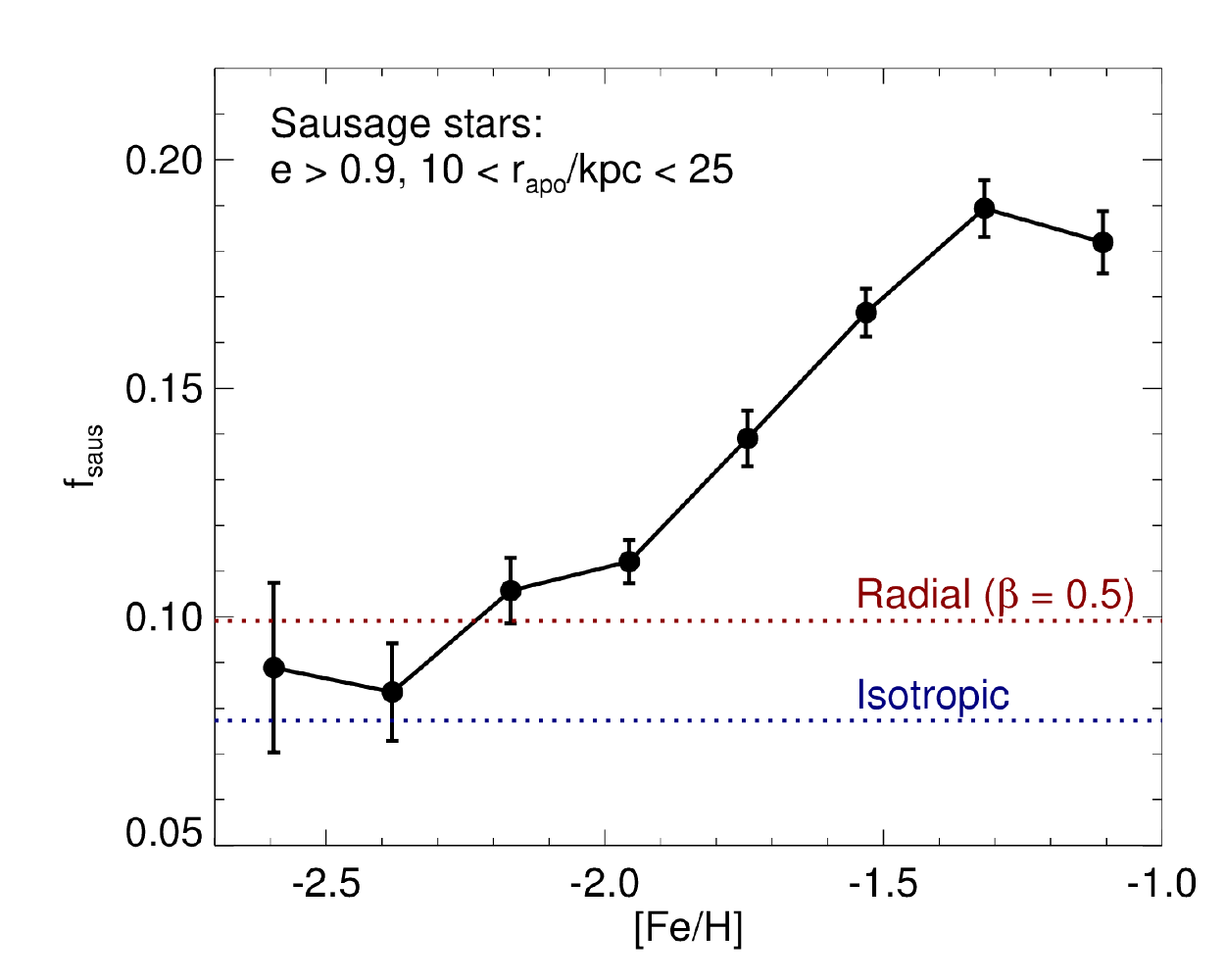}
  \end{minipage}
  \hspace{-15pt}
  \begin{minipage}{0.4\linewidth}
    \centering
   \includegraphics[width=8cm, height=6.4cm]{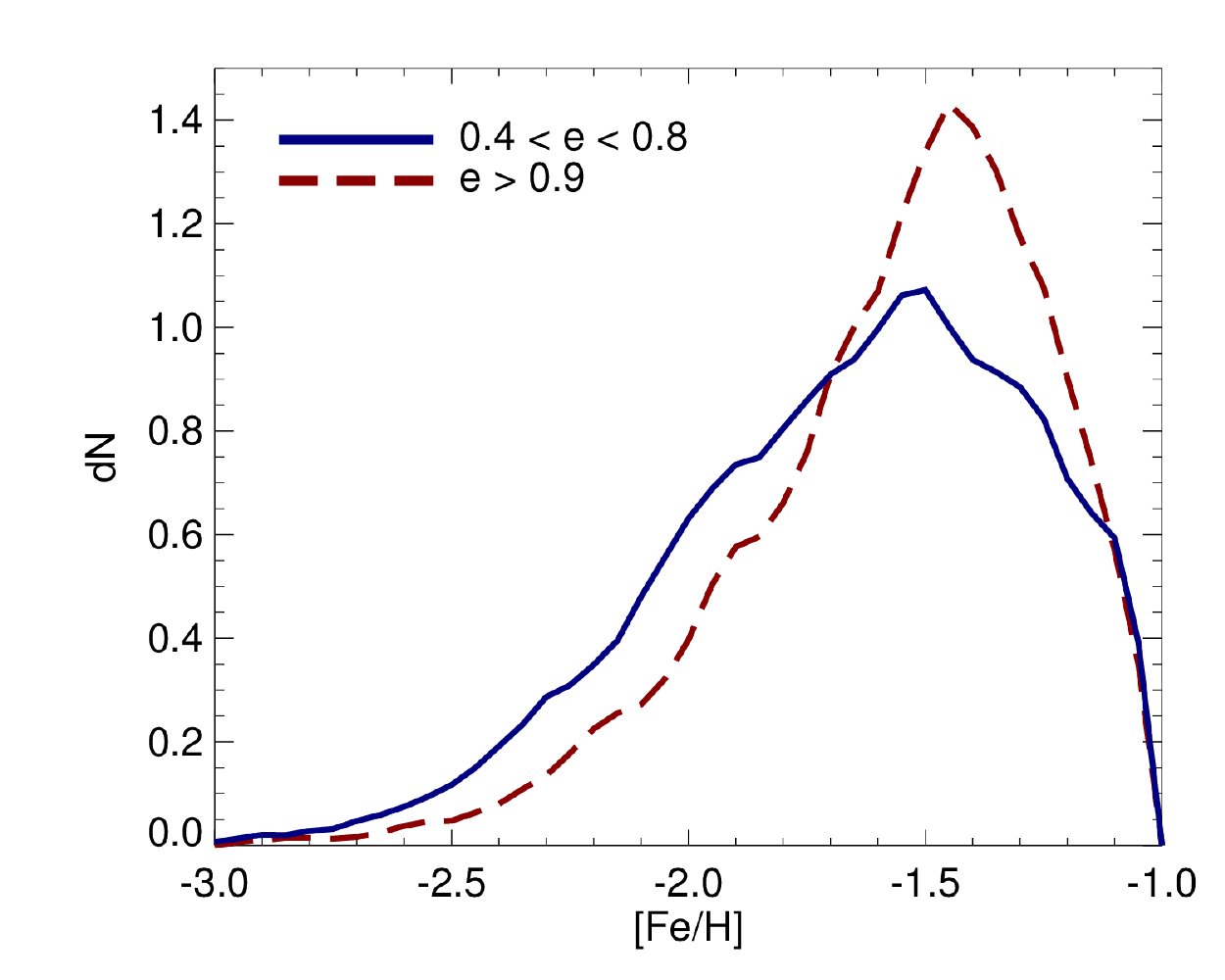}
  \end{minipage}
    \caption{\textit{Left panel:} The fraction of main sequence stars on radial orbits ($e > 0.9$) with apocenters in the range $ 10 < r_{\rm apo}/\mathrm{kpc} < 25$ as a function of metallicity. Here, we only consider stars with eccentricity $e > 0.5$ in order to exclude disk stars. The navy blue and dark red dotted lines gives the fractions for toy models with isotropic and radial ($\beta=0.5$) orbits, respectively. The ``sausage stars" are discernible down to  [Fe/H] $\sim -2$, but they become more evident at higher metallicity. \textit{Right panel:} The metallicity distribution of high eccentricity ($e > 0.9$) stars (dashed red line). For comparison, we show the distribution for stars with eccentricity $0.4 < e < 0.8$  (solid blue line). The ``sausage'' progenitor has higher median metallicity than the average halo population ([Fe/H]  $=-1.5$), and is likely at least as massive as the Sagittarius dSph ($M_{\rm star} \gtrsim10^8 M_\odot$).}
    \label{fig:fsaus_met}
\end{figure*}

In the left panel of Fig. \ref{fig:fsaus_met} we show the fraction of ``sausage" stars as a
function of metallicity. Here, we only use the main sequence sample,
which has larger numbers and more reliable metallicity
measurements. To select stars in this very radial component, we pick
stars with high eccentricity ($e > 0.9$) and with apocenters in the range $10 < r_{\rm
  apo}/\mathrm{kpc} < 25$. When calculating the fractions, we only consider stars with eccentricity $e > 0.5$ in order to minimize the contribution from disk stars. Thus, we define the fraction of ``sausage" stars as:
  \begin{equation}
 f_{\rm saus} = \frac{N(e > 0.9, 10 < r_{\rm apo}/\mathrm{kpc} < 25)}{N(e > 0.5)} 
 \end{equation}
 Note that these fractions should not be
taken as absolute as we are likely including field halo stars in the
selection and, moreover, we are excluding stars belonging to the same
progenitor with slightly different eccentricity. For comparison, we
show these fractions for toy models with the same spatial
distribution as the main sequence sample, but with an isotropic
($\sigma_\phi=\sigma_\theta = \sigma_r=120$ km s$^{-1}$) and
radially-biased ($\beta = 1-\sigma^2_{\rm tan}/\sigma^2_r = 0.5$)
velocity ellipsoid. 

The right panel of Fig. \ref{fig:fsaus_met} shows the metallicity distribution for 
high eccentricity stars ($e > 0.9$), with apocenters in the range $10 < r_{\rm
  apo}/\mathrm{kpc} < 25$. For comparison, the distribution for stars with $0.4 < e < 0.8$ is shown
  with the solid blue line.

Fig. \ref{fig:fsaus_met} shows that the stars
belonging to the ``sausage'' are discernible down to [Fe/H] $\sim -2$,
but they dominate at higher metallicity. This increase with
metallicity indicates that the progenitor has higher [Fe/H] than the
average halo ([Fe/H\ $\sim -1.5$, e.g. \citealt{an13}), which implies
  that it is at least as massive as the Fornax or Sagittarius dwarf
  spheroidals \citep[$M_{\rm star} \gtrsim10^8 M_\odot$ see
    e.g.,][]{kirby13, gibbons17}.

\section{Discussion and Conclusions}

We have used \textit{Gaia} DR2 proper motions and SDSS spectroscopy to
measure orbital properties of stars in the stellar halo. In
particular, we focus on the apocenter and pericenter distributions of
the halo stars. We find that both local samples of main sequence stars
and more distant samples of BHB stars have a relatively metal-rich
component on very radial ($e \sim 0.9$) orbits. Moreover, this
``sausage-like" component has apocenters that coincide with the
measured break radius of the Milky Way stellar halo.

The break radius in the Milky Way halo, beyond which the halo star
counts fall off significantly more rapidly, could relate to a
transition between two halo populations with different origins
\citep[e.g.,][]{carollo07,carollo10}. However, in \cite{deason13} we
argued that this broken profile signifies the build-up of stars at
apocenters, potentially deposited by a group of dwarfs at similar
times or by one massive dwarf. This latter scenario has recently
gained traction. In particular, \cite{belokurov18b} found a strongly
radially biased metal-rich population in nearby main sequence stars,
which they argue has been deposited by a massive dwarf galaxy. Here,
we show that not only is this ``sausage" population present in more
distant halo samples, but their apocenters directly coincide with the
stellar halo break radius. Thus, thanks to the exquisite proper
motions provided by \textit{Gaia}, we are able to, for the first time,
directly show that the break radius is indeed the location of an
apocenter pile-up. The high metallicity of this accretion event, and
the influence of its apocenter on the stellar halo density profile,
suggests that we have detected the most dominant progenitor of the
inner stellar halo.

\acknowledgments
We thank an anonymous referee for providing useful comments on our paper.

A.D. is supported by a Royal Society University Research
Fellowship. A.D. also acknowledges the support from the STFC grant
ST/P000541/1. The research leading to these results has received
funding from the European Research Council under the European Union's
Seventh Framework Programme (FP/2007-2013) / ERC Grant Agreement
n. 308024.

 This work has made use of data from the European Space
Agency (ESA) mission {\it Gaia}
(\url{https://www.cosmos.esa.int/gaia}), processed by the {\it Gaia}
Data Processing and Analysis Consortium (DPAC,
\url{https://www.cosmos.esa.int/web/gaia/dpac/consortium}). Funding
for the DPAC has been provided by national institutions, in particular
the institutions participating in the {\it Gaia} Multilateral
Agreement.

\bibliographystyle{aasjournal}
\bibliography{mybib}

\end{document}